# Intercell Moiré Exciton Complexes in Electron Lattices

Authors: Xi Wang[1,2*], Xiaowei Zhang[3*], Jiayi Zhu[1*], Heonjoon Park[1*], Yingqi Wang[1], Chong Wang[3], William Holtzmann, Takashi Taniguchi[4], Kenji Watanabe[5], Jiaqiang Yan[6], Daniel R. Gamelin[2], Wang Yao[7,8#], Di Xiao[3,1,9#], Ting Cao[3#], and Xiaodong Xu[1,3#]

[1]Department of Physics, University of Washington, Seattle, WA, USA
[2]Department of Chemistry, University of Washington, Seattle, WA, USA
[3]Department of Materials Science and Engineering, University of Washington, Seattle, WA, USA
[4]International Center for Materials Nanoarchitectonics, National Institute for Materials Science, Tsukuba, Ibaraki 305-0044, Japan
[5]Research Center for Functional Materials, National Institute for Materials Science, Tsukuba, Ibaraki 305-0044, Japan
[6]Materials Science and Technology Division, Oak Ridge National Laboratory, Oak Ridge, Tennessee, 37831, USA
[7]Department of Physics, University of Hong Kong, Hong Kong, China
[8]HKU-UCAS Joint Institute of Theoretical and Computational Physics at Hong Kong, China
[9]Pacific Northwest National Laboratory, Richland, Washington, United States

[*]These authors contribute equally to this work.
[#]Correspondence to: xuxd@uw.edu; tingcao@uw.edu; dixiao@uw.edu; wangyao@hku.hk

**Abstract**: **Excitons, Coulomb-bound electron-hole pairs, play a fundamental role in both optical excitation and correlated phenomena in solids. When an exciton interacts with other quasi-particles, few- and many-body excited states, such as trions, exciton Fermi-polarons, Mahan excitons can appear. Here, we report a new interaction between exciton and charges enabled by unusual quantum confinement in 2D moiré superlattices, which results in novel exciton many-body ground states composed of moiré excitons and correlated electron lattices. Unique to H-stacked (or 60°-twisted) WS$_2$/WSe$_2$ heterobilayer, we found that the interlayer atomic registry and moiré structural reconstruction leads to an interlayer moiré exciton (IME) whose hole in one layer is surrounded by its partner electron's wavefunction spread among three adjacent moiré traps in the other layer. This 3D excitonic structure can enable large in-plane electrical quadrupole moments in addition to the vertical dipole. Upon doping, the electric quadrupole facilitates the binding of IME to the charges in neighboring moiré cells, forming an intercell charged exciton complex. The exciton complex is unveiled by the IME photoluminescence energy jumps when the electron lattices form at both fractional and integer-filled moiré minibands, with replica-like spectral features between successive integer moiré fillings. Our work provides the framework in understanding and engineering emergent exciton many-body states in correlated moiré charge orders.**

**Main text**

In two-dimensional semiconducting moiré superlattices, such as MoSe$_2$/WSe$_2$ and WS$_2$/WSe$_2$, the superlattice potential results in arrays of trapping sites, giving rise to a new class of elementary photoexcitation - moiré excitons[1-5]. Due to the type-II band alignment of the host heterostructure, the ground-state moiré excitons consist of electrons and holes localized in opposite layers[6]. The interlayer moiré excitons possess unique physical properties, such as layer-stacking dependent optical selection rules and Landé-g factors[1], coupled spin-valley pseudospin degrees of freedom[7,8], arrays of interacting single emitters[2,9], and synthetic gauge fields for possible topological excitonic lattice[9]. On the other hand, semiconducting moiré superlattices are also powerful laboratories for exploring tunable many-body electronic ground states due to exceptionally strong Coulomb interactions[10-22]. Therefore, these moiré superlattices offer an opportunity to explore the interplay of moiré excitons with a correlated charge background.

Interlayer moiré excitons have been used to sense correlated charge states. A simple correlation between the enhancement of photoluminescence (PL) and the formation of both integer and fractionally filled moiré minibands has been well established [14,15,20,23]. However, little is known about how interlayer moiré excitons interplay with the variety of charge orders, and thus the resulting correlated states of the interacting exciton-electron lattice system. For instance, is the interaction between exciton and moiré trapped charges similar to the simple trion picture, i.e., a neutral exciton bound to an extra charge? Or is there a new type of many-body excited states formed as the moiré excitons are embedded in the electron lattice? Understanding these fundamental questions will provide a framework for the burgeoning field of interacting opto-moiré quantum matter[10-20] as well as the field of strongly correlated phenomena in optical excitations[24].

Here, we discover a new class of ground-state moiré excitons that can simultaneously possess a large in-plane quadrupole and out-of-plane dipole moment, a 3D structure unique to the H-stacked heterobilayer. By sequentially charging the moiré traps, the developing in-plane quadrupole moment enables the binding of such a moiré exciton to the ordered charges localized in the neighboring moiré cells, forming an intercell charged moiré exciton complex – a new type of many-body excitonic state. Distinct from trion[25-27] and exciton-Fermi polaron physics[28], the intercell moiré exciton complex manifests as replica of the exciton photoluminescence (PL) spectral features as a function of doping. We find tens of meV blueshift in the moiré exciton PL peaks across the integer moiré filling, resulting from intracell Coulomb repulsion, and a subsequent 5 to 10 meV redshift as the moiré minibands are fractionally filled. The latter value corresponds to the binding energy of the intercell moiré exciton complex.

**Moiré spatial charge distribution**

The moiré electronic structure sensitively depends on layer stacking and local interlayer atomic registry. We performed density-functional theory calculations to capture such dependence (Methods). Figures 1a and b depict R- and H-stacked moiré supercell, respectively, with three high-symmetry positions marked as **A**, **B**, and **C**. Using R stacking as an example, **A**, **B**, and **C** regions have local stacking arrangements in $R_h^h$, $R_h^M$, and $R_h^X$, respectively. Here, the notation $R_h^\mu$ represents the μ (μ=h (hollow), M (W), X (S)) site of the electron layer (WS$_2$) vertically aligned with the hexagonal center h of the bottom hole layer WSe$_2$. To capture the significant lattice reconstruction in the moiré superlattices[29], we performed structural relaxation calculations and found different strain distribution between R- and H-stacked heterobilayers (Extended Data Fig. 1). This distinct feature is due to the different structural energy distribution of local stacking (Extended Data Fig. 2).

Based on these reconstructed moiré structures, we calculate the spatial wavefunction of the valley-moiré conduction band bottom and valence band top, corresponding to the expected positions of the frontier electron- and hole-orbitals, respectively. For R stacking, both the electron and hole are localized at **C** (Fig. 1c). Therefore, for the ground-state moiré exciton, the electron in the $WS_2$ layer is vertically aligned with the hole in the $WSe_2$ layer, i.e., no lateral misalignment between the electron and hole wavefunctions. The moiré exciton ground state is thus expected to only have an appreciable out-of-plane electric dipole.

In contrast, the wavefunction in the H-stacked heterobilayer has very different spatial distributions. While the hole is largely localized at **C**, the electron is mostly localized at a different region, **A**. For the ground-state moiré exciton, the electron and hole wavefunction are thus expected to be vertically mis-aligned. For a fixed hole at **C** in $WSe_2$, the bound electron wavefunction is expected to spread over three adjacent **A** positions of the $WS_2$ layer with a 3-fold rotational symmetry (Fig. 1d). As we will discuss next, this exciton wavefunction features both in-plane electric quadrupole and out-of-plane dipole moments, which lead to the formation of a new class of intercell moiré exciton complex in the presence of the electron lattice (i.e., correlated charge order).

**Layer stacking dependent photoluminescence**

In our experiment, we fabricated both R- and H-stacked $WS_2/WSe_2$ heterobilayers (See methods). Figures 2a-b show optical, atomic force, and piezoresponse force microscopy images of a representative H-stacked sample with 7.6 nm moiré wavelength. We performed photoluminescence (PL) measurements of interlayer moiré excitons. The optical excitation power is 50 nW with the excitation energy specified in figure captions. Figure 2c shows the PL spectrum of an R-stacked heterobilayer at charge neutral (Device-R1) at 10 K. A pronounced interlayer exciton PL peak is observed at ~1.37 eV, consistent with previous reports[3,14,15].

We then measured interlayer moiré exciton PL as a function of doping with the electric field fixed at zero. Figure 2d shows the results from Device-R1 with the filling factor $v$ as indicated (See Methods and Extended Data Fig. 3 for the assignment of $v$). The data are plotted in log scale to highlight the weak features for $|v|$ above 1. Linecuts at select filling factors are shown in Extended Data Fig. 4. The PL exhibits sharp intensity jumps as doping reaches both fractional and integer filled moiré minibands, due to the formation of charge ordered states. Near each fractional filling factors, interlayer exciton PL intensity plot as a function of doping and emission energy shows sharp tilted line features. Since the peak energy along the titled line always blue shifts for either increasing electron or hole doping, this rules out the possible stark effect as its cause due to unbalanced top and bottom gates. Instead, we speculate that the charge ordered states are robust even with a small number of excessive electrons. The excessive electrons can act as a screening cloud until the number of electrons meets the requirement to form another correlated insulating state at a new fractional filling. The precise slope of the diagonal features is determined by the filling factor dependent local dielectric environment of each device.

We extract the PL peak energies at those most appreciable charge ordered states and plot them versus $v$ in Fig. 2e. The dashed lines represent the peak energies at all measured gate voltages.

Among the fractional *v* states between two consecutive integer *v*, the energy of PL peaks varies little (Fig. 2d-e). As |*v*| goes across an integer value, the peaks blueshift. For instance, the blueshift is about 5 and 17 meV, respectively, as *v* crosses -1 and 1 (corresponding to the charge ordered states with one hole and one electron per moiré unit cell). Extended Data Figs. 5 and 6 show similar measurements of additional R-stacked heterobilayer devices. Although some details differ (such as the number of observed correlated charge orders), likely due to sample-to-sample variations, the dependence of the PL peak energy as a function of *v* is consistent. The observation is distinct from that in $MoSe_2/WSe_2$ heterobilayer, where moiré exciton PL peaks have a discrete redshift by about 7 meV upon formation of moiré trions by doping[25-27].

In contrast, H-stacked samples show drastically different doping dependence of the exciton PL. Figure 2f shows the PL intensity plot with integer filling factors indicated (Extended Data Fig. 3). The extracted PL peak energies of prominent charge ordered states are plotted in Fig. 2g. Over two dozen correlated states at both integer and fractional moiré fillings are observed. Despite similar jumps of PL intensity at these fillings, the dependence of peak energy on *v* is distinct in H-stacked samples compared to the R-stacked ones. Starting from charge neutrality, the first charge order appears at fractional moiré filling *v*=±1/7. This is evident by the enhanced PL intensity and sudden peak redshifts of about 5 meV (*v*=1/7) and 7 meV (*v*=-1/7) (Extended Data Fig. 7). Here, |*v*|=1/7 corresponds to a commensurate triangular charge lattice. When the next charge order at *v*=±1/3 forms, a further redshift of 8 meV (*v*=1/3) and 7 meV (*v*=-1/3) are observed. These discrete redshifts are missing in all R-stacked samples in the same doping range. Upon further doping, PL peak energies have little variation until reaching v=±1, where the PL peaks exhibit sharp energy jumps by +22 and +42 meV for the *v*=-1 and +1 states, respectively. Similar to the peak redshift going from charge neutrality to |*v*|=1/7, the PL peak redshifts again when |*v*| is slightly larger than 1 to form the next fractionally filled states. This redshift is most pronounced from v=-1 to -8/7 of about 4 meV. The general spectral feature between $1 \leq |v| < 2$ resembles that between $0 \leq |v| < 1$ despite fine differences, such as the number of observed correlated states.

When $|v|$ exceeds 2, the spectral dependence on *v* is most appreciable on the hole side. The PL peak first blueshifts for another 18 meV at *v*=-2, and then exhibits discrete redshifts of about 4 meV as doping increases slightly. Upon further doping, the peak continues to blueshift. This behavior for *v* over -2 again resembles those of $0 \leq |v| < 1$, except that the signatures of correlated states are not clear likely due to effective screening of long-range Coulomb interactions at high doping. Extended Data Fig. 8 shows similar measurements of an additional H-stacked heterobilayer device. Note that the R and H-stacked samples in Extended Data Figs. 6 and 8 are from the different parts of the same device, i.e., fabricated simultaneously. This further confirms the observation is due to stacking, rather than from device fabrication variation.

**Intercell moiré exciton complex**

The above experimental observations unveil the stacking dependent interactions between interlayer excitons and electron lattice in moiré heterostructures. Figure 3a illustrates the key idea. Upon doping, moiré sites are fractionally filled. To avoid double occupancy and onsite Coulomb interactions, the interlayer exciton will only occupy an empty moiré site. As presented earlier, for R-stacked heterobilayer, the electron and hole of the interlayer moiré exciton are vertically aligned, leading to a weakly repulsive interaction between the excitonic dipole and its adjacent moiré trapped charge (Fig. 3a, top panel). Thus, this interlayer exciton does not bind to the moiré charges. The fractional insulating states largely behave as an inert background and the PL peak energy experiences little variation with an estimated blueshift of < 2 meV until integer filling (See Methods and Extended Data Fig. 9).

In contrast, a large electric quadrupole can form for interlayer moiré excitons in the H-stacked heterobilayer (Fig. 3a, bottom panel), allowing the exciton to bind to the electron lattice. At $v$=-1/7, the dilute hole lattice regime, the interlayer exciton in an empty moiré cell can bind with the hole in the adjacent cell (Fig. 3b). This nearest-neighbor intercell Coulomb binding results in the observed ~ 7 meV redshift of PL peaks, consistent with the theoretical estimation (Extended Data Fig. 9). As the hole density increases, the interlayer exciton then interacts with the entire electron lattice, forming a charged intercell moiré exciton complex with an additional redshift. Figure 3c illustrates the intercell moiré exciton complex at $v$=-1/3. The observed PL peak energy difference between $|v|$=1/3 and charge neutral thus corresponds to a binding energy of the intercell moiré exciton complex of about 14 meV, consistent with the estimation of 15 meV from theory (Extended Data Fig. 9).

With one hole in each moiré unit cell ($v$=-1), the interlayer moiré exciton interacts with an intra-moiré cell hole (Fig. 3d). The interaction involves both intracell hole-hole repulsion and electron-hole attraction, which have opposite signs. Since the electron and hole wavefunction are laterally misaligned in H stacked heterobilayer, it has a larger electron-hole separation than that in the R stacked one with vertically aligned electron-hole wavefunctions. Therefore, the excitonic Coulomb binding in the former is weaker than in the latter. Upon formation of a sizable charge gap at integer filling, this difference leads to a much larger increase of the PL peak energy (~22 meV at $v$ = -1) in the H-stacked sample than that in the R-stacked heterobilayer (~5 meV). Upon further doping, the interlayer exciton will then again bind to the additional hole in the adjacent moiré site, forming a charged intercell moiré exciton. The above-described process repeats itself at either electron or hole integer moiré filling. Note that the energy jump at $v$ = 1 is about 42 meV, larger than that for the $v$ = -1 state. This asymmetry between the electron lattice and hole lattice may arise from dissimilar electron and hole wavefunctions of the interlayer exciton, which leads to different intracell Coulomb interaction strength.

**Optical selection rules**

We find layer stacking dependent PL polarization of the interlayer moiré exciton. Figures 4a and b show the circular polarization resolved photoluminescence of the R-stacked heterobilayer (Device-R1) and the associated degree of polarization as a function of doping. The optical excitation energy is 1.678 eV, in resonance with $WSe_2$ 1s exciton. PL is co-circularly polarized in general, except cross circularly polarized for $v$ beyond -1/2. Extended Data Fig. 10 shows similar measurement but with an 8 T out-of-plane magnetic field. The cross-polarized feature for v <-1 remains, implying the selection rules is determined by the symmetry of the local atomic registry. For the states of $v$ between +1 and -1, the degree of circular polarization is weak, while for $v$ above 1, PL is strongly co-circularly polarized. In contrast, the H-stacked heterobilayer possesses distinct optical selection rules. Figure 4c shows the circular polarization resolved PL measurements as a function of doping (taken at a different spot and cool down cycle compared to the data in Fig. 2.). The corresponding degree of circular polarization is shown in Fig. 4d. The charge-neutral interlayer moiré exciton and the intercell moiré exciton complex are both cross circularly polarized, although the degree of polarization is weak for $|v|$ <1. Consistent polarization resolved measurements from additional R and H stacked samples are presented in Extended Data Figures 6 and 8, respectively.

The distinct doping dependence of PL polarization is ascribed to the different atomic registries between the R- and H-stacked heterobilayer. The optical selection rule for interlayer electron-hole recombination is determined by both their spin-valley indices and the local stacking registry[30]. As shown in Fig. 4e, for the R-stacking, the electron and hole recombine at $R_h^X$ regardless of doping, and the K valley of WSe$_2$ is nearly momentum-aligned with the K valley of WS$_2$. After σ+ pump, the photo-generated electron at the K-valley of WSe$_2$ can relax into the K and -K valleys through spin-flip and valley-flip scattering. Previous works have shown that the relaxation rates for these two paths are similar[27,31]. In both charge-neutral and electron-doped cases, since the hole predominately populates at the WSe$_2$ K valley, the emission helicity is σ+. In the hole-doped case, the hole has a comparable population at WSe$_2$ ±K valleys. In addition, noting that the population of photo-generated electrons at the WS$_2$ -K valley is likely to dominate over the K valley population, the polarization is reversed at certain hole-doping level.

For H-stacking, due to their lateral separation, the electron and hole can recombine at $H_h^h$ and/or $H_h^X$ depending on doping as shown in Fig. 4f. Additionally, the WSe$_2$ K valley is nearly momentum-aligned with the WS$_2$ -K valley. Therefore, both the spin-valley index and recombination site affect the doping-dependence of PL polarization. In the charge-neutral case, the observed weak polarization suggests that the recombination rates at $H_h^h$ and $H_h^X$ are likely comparable to each other. In the electron-doped case, the exciton prefers to distribute around the electron ($H_h^h$ site) to form bound exciton-electron-lattice many-body states, resulting in cross-polarization. In the hole-doped case, similar to R-stacking, the population of photo-generated electrons at K valley of WS$_2$ dominates over the -K valley population. In this case, the exciton prefers to distribute around the hole ($H_h^X$ site) to maximize the attractive exciton-hole-lattice interaction, yielding cross-polarization. Overall, these polarization dependent results are consistent with our understanding of the unique interlayer moiré exciton and their charged intercell complexes in H-stacked heterobilayers.

**Summary**

Our work reveals a unique stacking dependent interlayer moiré exciton wave function, which leads to repulsive and attractive exciton - moiré carrier interactions in R- and H-stacked heterobilayers, respectively. This provides a framework in understanding the interlayer moiré exciton properties, such as luminescence energy and valley polarization, in the presence of multitudinous correlated charge orders. The in-plane electrical quadruple moment of the interlayer exciton in H-stacked sample enables its binding with the electron lattice to form a new type of exciton many-body ground states, with observed binding energy ∼ 14 meV at 1/3 hole filling consistent with our theory. These findings will facilitate future work in unraveling and manipulating light-matter interactions in the rapidly developing field of semiconducting moiré quantum matters, which emerged as a powerful platform to study correlated and topological phenomena with unprecedented tunability[10-22,32-38].

**Methods:**

**Sample fabrication.**

Mechanically exfoliated monolayers of WS$_2$ and WSe$_2$ were stacked using the dry-transfer technique. Bulk WSe$_2$ crystals were lab grown and bulk WS$_2$ crystals were purchased from HQ Graphene. The crystal orientation of the individual monolayers was first determined by linear-polarization resolved second-harmonic generation before transfer. All the heterostructures are carefully AFM cleaned with minimal force to obtain homogeneous sample area. The layer twist angle was determined by piezoresponse force microscopy (PFM) before encapsulating with hexagonal boron nitride and graphite.

**Optical measurements.**
PL and differential reflectance measurements were performed using a home-built confocal microscope in reflection geometry. The sample was mounted in a close-cycled cryostat with temperature kept at either 5 K or 10 K, unless otherwise specified. He-Ne laser (1.96 eV) was used in unpolarized PL measurements. A power-stabilized and frequency-tunable narrow-band continuous-wave Ti:sapphire laser (M$^2$ SolsTiS) resonant with the WSe$_2$ 1s intralayer exciton was used to excite the sample for polarization-resolved PL, with the combination of quarter-wave plates, half-wave plates, and linear polarizers. The excitation power for PL measurement is 50-150 nW. For reflection spectroscopy, a halogen lamp was used as a white light source. The output of white light was passed through a single-mode fiber and collimated with a triplet collimator. The beam was then focused onto the sample with a 40x long working distance objective (NA=0.6). The excitation power is below 5 nW. Reflectance and PL signals were dispersed by a diffraction grating (600 grooves per mm) and detected on a silicon CCD camera. The PL was spectrally filtered from the laser using a long-pass filter before being directed into a spectrometer.

**Estimation of filling factor based on doping density.**
The BN thickness of Device R1 are 42 nm (top) and 43 nm (bottom). The BN thickness of Device H1 are 16 nm for both top and bottom gates. The doping density was calculated as $C_t\Delta V_t + C_b\Delta V_b$ based on the parallel-plate capacitance model, where $C_t$ and $C_b$ are the capacitance of the top and bottom gates, and $\Delta V_t$, $\Delta V_b$ are the applied gate voltages. $\varepsilon_{hBN} \sim 3$ was used as the dielectric constant of BN in the calculation. The moiré lattice constant was determined by PFM. Doping density and moiré lattice constant were used to estimate the filling factors initially. The filling factors were then compared and adjusted with the assignment of integer filling factors based on optical reflectance measurement (dR/R) and PL.

**First-principles calculations of moiré structures and electronic structures**
First-principles calculations were performed using density functional theory with the Perdew-Burke-Ernzerhof functional[39] as implemented in the SIESTA package[40]. The heterobilayer was constructed using 26×26 WS$_2$ and 25 ×25 Wse$_2$ supercells, where the lattice constants of WS$_2$ and Wse$_2$ were taken to be their optimized values, 3.17 Å and 3.30 Å, respectively (within < 0.5% error). A supercell arrangement was used, with the out-of-plane axis set to 21 Å to avoid interactions between the moiré heterobilayer and its periodic images. Optimized norm-conserving Vanderbilt pseudopotentials were used[41]. In the relaxation of the structures, single-zeta basis was chosen and dispersion corrections within the D2 formalism were used to include the van der Waals interactions[42]. The structure was fully relaxed until the force on each atom was smaller than 0.03 V/Å. The normal strain distribution was extracted from the displacement field of W atoms and chalcogen atoms with respect to the unrelaxed heterobilayer. For the calculations of W-atom strain distribution, the W-W distance was obtained by averaging the distance along six nearest-neighbor W atoms. In the calculations of electronic structure, we used the double-zeta plus polarization basis and spin-orbit coupling.

**Modelling exciton-charge-lattice interactions**

In the case of small doping density, the exciton-hole-lattice interaction can be approximately treated using screened Coulomb interactions. Thus, the exciton energy shift upon doping relative to the neutral exciton energy can be modeled as the sum of two terms, i.e., the attractive interaction between the excitonic electron and the hole lattice, and the repulsive interaction between the excitonic hole and the hole lattice. The potential $V_c$ from the hole lattice acts on the excitonic electron or excitonic hole, having a form

$$V_c(\boldsymbol{r}) = \sum_{i,\boldsymbol{R}} \frac{q_i}{4\pi\varepsilon_0\varepsilon_r|\boldsymbol{r} - \boldsymbol{r}_i - \boldsymbol{R}|},$$

Where $q_i$ is the charge of the doped hole, $\boldsymbol{r}_i$ is the position of the doped hole in the primitive unit cell of the hole lattice, and $\boldsymbol{R}$ is the hole-lattice vector (see Fig. 3). As the smallest distance in the denominator of the summation is ~ 5 nm, much larger than the thickness of the WS$_2$-Wse$_2$ heterostructure, $\varepsilon_r$ can be taken as the relative permittivity of the environment[37], i.e., the encapsulating h-BN with $\varepsilon_r$ ~ 4.0. In order to deal with the long-range nature of Coulomb interaction, we adopt the three-dimensional Ewald summation method to rewrite $V_c(\boldsymbol{r})$ as,

$$V_c(\boldsymbol{r}) = \frac{4\pi}{\Omega}\sum_{i,\boldsymbol{G}\neq 0} q_i \frac{e^{-G^2/4\alpha^2}}{G^2} e^{-i\boldsymbol{G}\cdot(\boldsymbol{r}-\boldsymbol{r}_i)} + \sum_{i,\boldsymbol{R}} q_i \frac{\text{erfc}(\alpha|\boldsymbol{r}-\boldsymbol{r}_i-\boldsymbol{R}|)}{|\boldsymbol{r}-\boldsymbol{r}_i-\boldsymbol{R}|}.$$

Here, **G** is the reciprocal lattice vector of the moiré lattice. To avoid the artificial interactions from the hole lattice images along surface normal, a supercell is constructed with large out-of-plane thickness equal to 5 times the lattice constant of the moiré lattice. The thickness has been tested and converged to ensure the Ewald summation is equivalent to two-dimensional form. The **G**=0 term is skipped in the summation due to compensating charge in the dual gates, which ensures overall charge neutrality. The parameter $\alpha$ is chosen as around 0.003 Bohr$^{-1}$ and the cut-off for the summation over **R** and **G** is carefully tested for moiré lattices with different filling factors. The doping-dependence of exciton energy shift is given in Extended Data Fig. 9 for R- and H-stacked heterobilayer. The energy shift of the electron-doping case is similar.

**Acknowledgements:** Research on the exciton many-body ground states is mainly supported by DoE BES under award DE-SC0018171. Measurements on the R-stacked moiré superlattice is supported as part of Programmable Quantum Materials, an Energy Frontier Research Center funded by the U.S. Department of Energy (DOE), Office of Science, Basic Energy Sciences (BES), under award DE-SC0019443. The first-principles calculation is mainly supported by NSF MRSEC DMR-1719797. Computational resources were provided by HYAK at the University of Washington. The theoretical analysis and modeling effort is supported by DOE DE-SC0012509. Device fabrication is partially supported by Army Research Office (ARO) Multidisciplinary University Research Initiative (MURI) program (grant no. W911NF-18-1-0431). The AFM-related measurements were performed using instrumentation supported by the U.S. National

Science Foundation through the UW Molecular Engineering Materials Center (MEM-C), a Materials Research Science and Engineering Center (DMR-1719797). WY acknowledges support by the University Grants Committee/Research Grant Council of Hong Kong SAR (AoE/P-701/20, HKU SRFS2122-7S05), and Tencent Foundation. Bulk $WSe_2$ crystal growth and characterization by JY is supported by the US Department of Energy, Office of Science, Basic Energy Sciences, Materials Sciences and Engineering Division. K.W. and T.T. acknowledge support from the Elemental Strategy Initiative conducted by the MEXT, Japan (Grant Number JPMXP0112101001) and JSPS KAKENHI (Grant Numbers 19H05790, 20H00354 and 21H05233). T.C. acknowledges support from the Micron Foundation. XX acknowledges support from the State of Washington funded Clean Energy Institute and from the Boeing Distinguished Professorship in Physics.

**Author contributions:** XX, WY, TC, DX conceived the project. XW, HP and JZ fabricated and characterized the samples. XW, JZ, HP and YW performed the measurements, assisted by WH. XW, XZ, CW, XX, WY, TC, DX, DRG analyzed and interpreted the results. XZ and CW performed density function calculations. TT and KW synthesized the hBN crystals. JY synthesized and characterized the bulk $WSe_2$ crystals. XX, XW, XZ, WY, TC, DX, DRG wrote the paper with input from all authors. All authors discussed the results.

**Competing Interests:** The authors declare no competing financial interests.

**Data Availability:** The datasets generated during and/or analyzed during this study are available from the corresponding author upon reasonable request.

**Figures:**

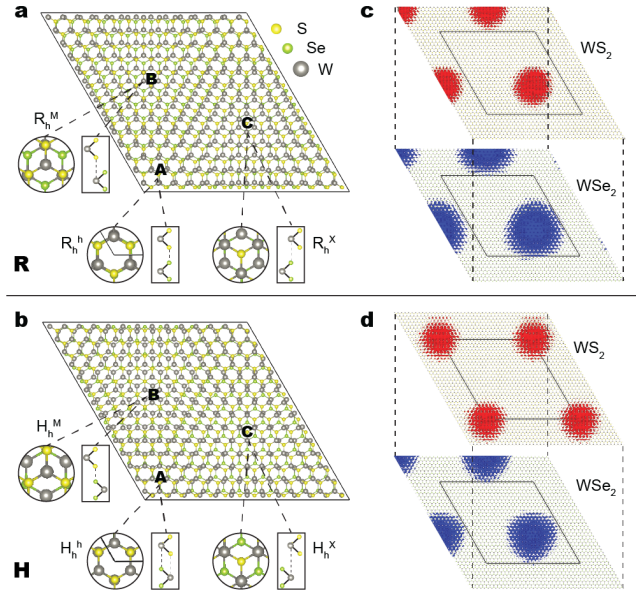

**Figure 1. Spatial charge distribution of interlayer moiré exciton. a**, Schematic of R-stacked WS$_2$/WSe$_2$ (top/bottom) heterobilayer. Zoom-in plots show the atomic registry of three high-symmetry positions in the moiré supercell, **A, B, C,** corresponding to local stacking arrangements $R_h^\mu$. Here $R_h^\mu$ represents the μ (μ=h (hollow), X (S), M (W)) site of the electron layer (WS$_2$) vertically aligned with h of the bottom hole layer (WSe$_2$). The true structure employed in the first-principles calculations has 3903 atoms per moiré supercell, larger than the cell size shown in (a) that exaggerates the moiré lattice mismatch for illustration. **b,** Similar plot as (a) but for H stacked heterobilayer. **c** and **d**, Calculated spatial Kohn-Sham wavefunction of the valley-moiré conduction band bottom and valence band top in (**c**) R- and (**d**) H-stacked heterobilayer, illustrating the expected spatial distributions of band-edge electrons and holes, respectively. Isosurface of squared wavefunction is taken at ~ 5% of its maximum value. For R stacking, both the electron and hole are localized at **C**. Therefore, for the ground-state moiré exciton, the electron in the WS$_2$ layer is vertically aligned with the hole in the WSe$_2$ layer (**c**). In contrast, in the H-stacked heterobilayer, while the hole is localized at **C**, the electron is mostly localized at **A**. For the ground-state moiré exciton with a fixed hole at **C** in WSe$_2$, the bound electron wavefunction largely spreads among the three adjacent **A** positions of the WS$_2$ layer (d). Therefore, in addition to the out-of-plane dipole, the interlayer moiré exciton has an in-plane electric quadrupole moment.

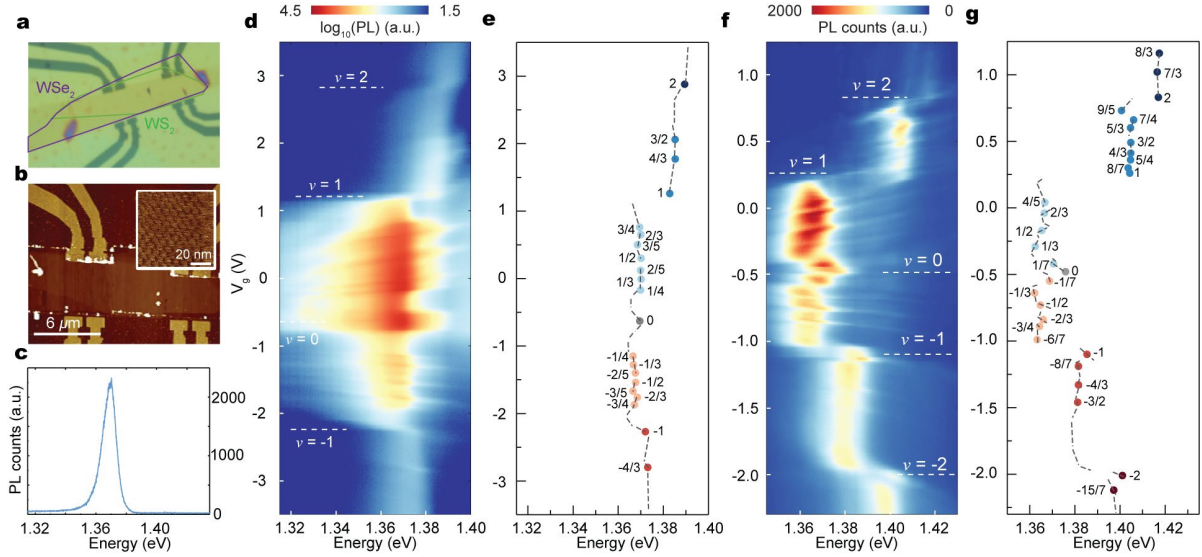

**Figure 2: Moiré exciton photoluminescence (PL) in the presence of correlated charge orders. a**, Optical microscope image of a near 60-degree twisted $WS_2/WSe_2$ heterobilayer. **b,** Corresponding atomic force microscope image. Inset: piezoresponse force microscopy image, showing moiré pattern with a triangular lattice. **c**, PL spectrum of an interlayer moiré exciton of an R-stacked device (Device-R1). **d**, Interlayer exciton PL intensity plot (log-scale) versus moiré filling factor $v$ of Device-R1. Optical excitation is 1.678 eV with the power 50 nW. **e**, Extracted PL peak energies (dots) for each integer and fractionally filled correlated charge states. The dashed lines indicate PL peak energies at all measured gate voltages. **f** and **g**, The same plots as (d) and (e) but for the H-stacked Device-H1. Optical excitation is 1.96 eV with the power 50 nW. See text for details.

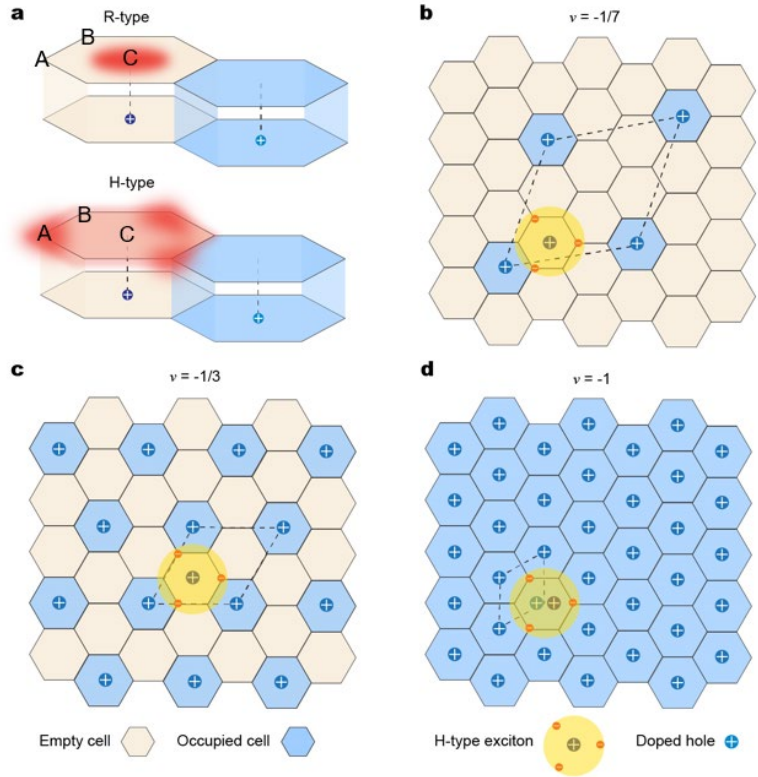

**Figure 3: Schematics of intercell moiré exciton complex. a,** Cartoons depicting the interlayer exciton and an additional hole trapped in the adjacent moiré unit cell. In R-stacking (top panel), since electron and hole wavefunctions are vertically aligned, the moiré interlayer exciton has a weak repulsive interaction with the hole. In contrast, the exciton in H-stacking (bottom panel) can have an attractive interaction with the hole by developing a large electric quadrupole that lies in-plane. **b-d,** Cartoons of intercell moiré exciton complexes for different hole filling factors in an H-stacked heterobilayer. The light yellow (empty site) and light blue (with one doped hole) hexagons denote the Wigner-Seitz cell of the moiré primitive unit cell. The dashed parallelogram indicates the primitive unit cell of the hole lattice. For $v = -1$, there exist intracell Coulomb interactions between the moiré exciton and doped hole.

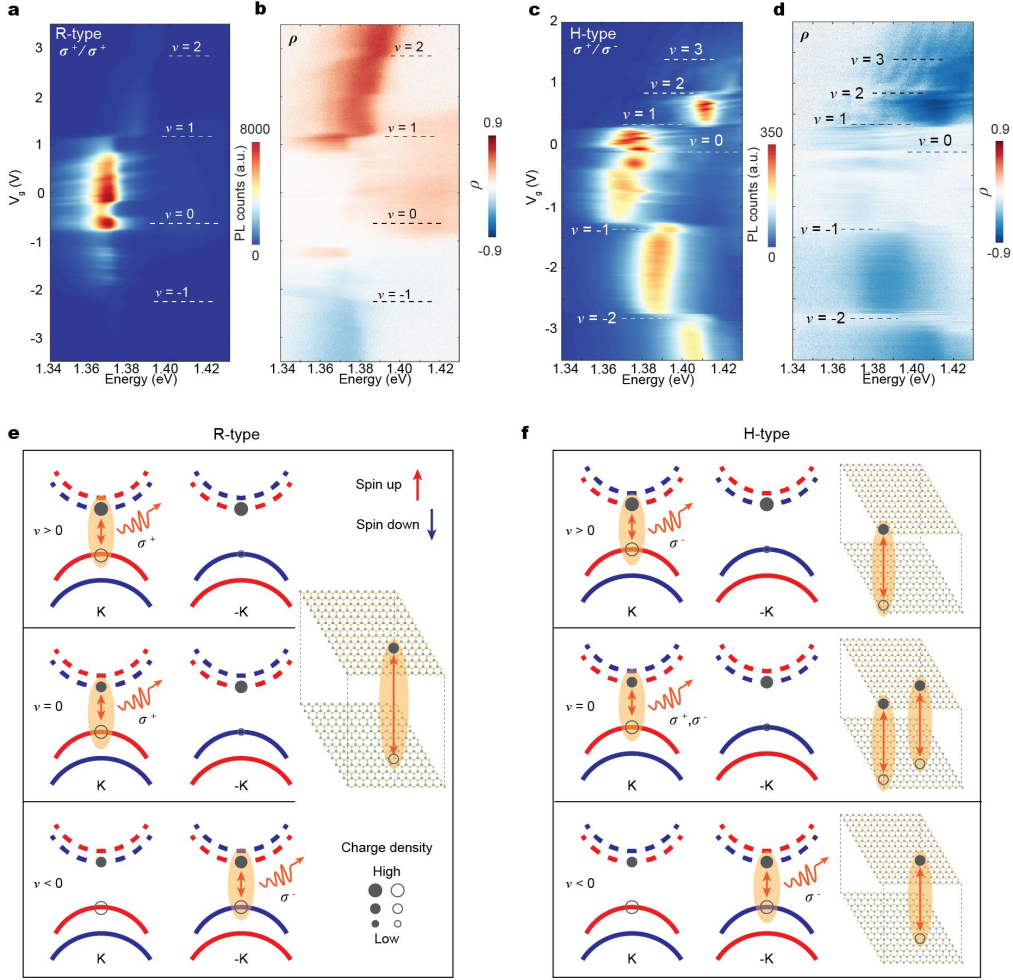

**Figure 4. Stacking dependent moiré exciton valley polarization. a**, Co-circularly polarized interlayer exciton PL for Device-R1 and **c**, Cross-circularly polarized interlayer exciton PL for Device-H1, under σ+ circularly polarized excitation. The excitation energy is in resonance with WSe$_2$ 1s exciton (~1.678 eV). The excitation power is 50 nW for R1 and 150 nW for H1. **b** and **d** plot the corresponding degree of circularly polarization $\rho = \frac{\sigma^+/\sigma^+ - \sigma^+/\sigma^-}{\sigma^+/\sigma^+ + \sigma^+/\sigma^-}$. **e** and **f,** Schematics of doping-dependent PL polarization for R- and H-stacked heterobilayers, respectively. Solid (open) circles denote electrons (holes) with the circle size indicating the population. Solid (dashed) parabolas represent valence (conduction) bands from WSe$_2$ (WS$_2$), with the red and blue color denoting spin up and down, respectively. The PL polarization is determined by the majority-carrier valley index and the location of recombination in the moiré cell. As indicated by the insets, the electron and hole recombine at $R_h^X$ for the R-stacking, while for H-stacking they recombine at $H_h^h$ and/or $H_h^X$ depending on the doping. See text for details.

# Extended Data for
# Intercell Moiré Exciton Complexes in Electron Lattices

Authors: Xi Wang[1,2*], Xiaowei Zhang[3*], Jiayi Zhu[1*], Heonjoon Park[1*], Yingqi Wang[1], Chong Wang[3], William Holtzmann[1], Takashi Taniguchi[4], Kenji Watanabe[5], Jiaqiang Yan[6], Daniel R. Gamelin[2], Wang Yao[7,8#], Di Xiao[3,1,9#], Ting Cao[3#], and Xiaodong Xu[1,3#]

Correspondence to: xuxd@uw.edu; tingcao@uw.edu; dixiao@uw.edu; wangyao@hku.hk

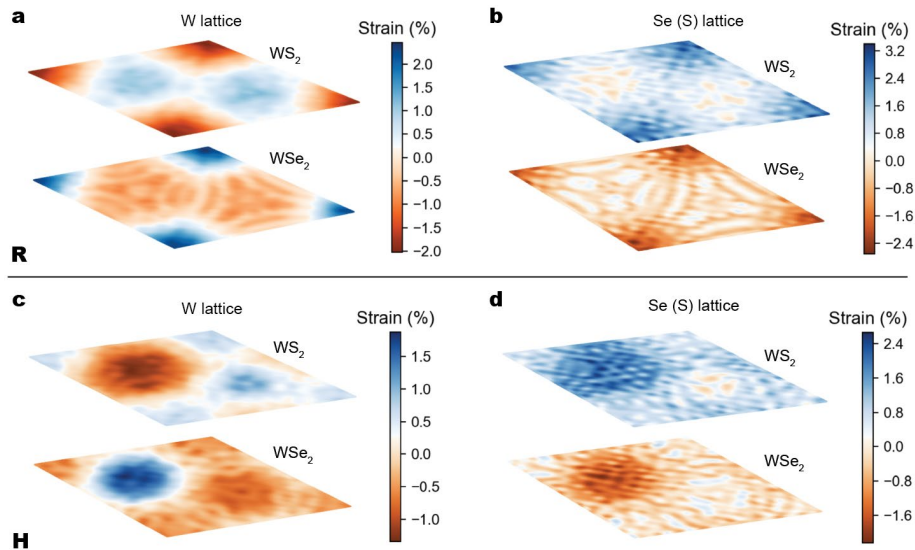

**Extended Data Figure 1. Calculated normal strain distribution of the R- and H-stacked heterobilayers. a,** and **c,** Calculated strain maps of the moiré unit cell for the fully relaxed W lattice in $WS_2$ and $WSe_2$ versus free-standing layers for (a) R- and (c) H-stacked heterolayers. **b,** and **d,** similar to **a** and **c**, but for the fully relaxed Se or S lattice in the moiré unit cell. The strain distributions are different in R- and H-stacked heterobilayers. Taking A site as the inversion center of the strain distribution as an example, the inversion symmetry weakly breaks in the former, but strongly breaks in the latter. This distinct feature is due to the different structural energy distribution of the local stacking (Extended Data Fig. 2).

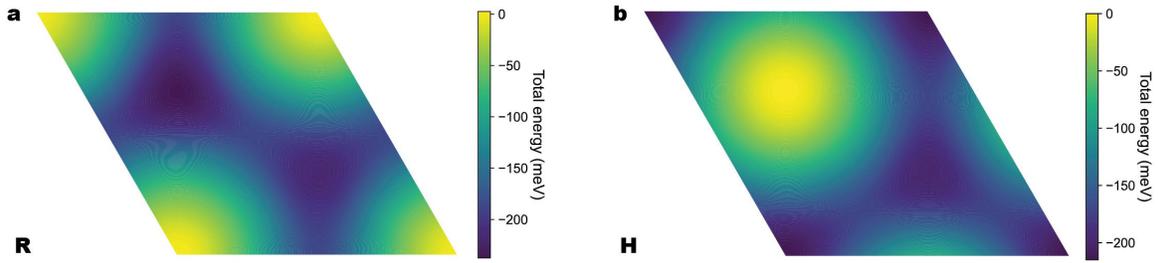

**Extended Data Figure 2. Calculated structural energy distribution for the local stacking configuration in the moiré cell for a,** R-stacked and **b,** H-stacked heterobilayers.

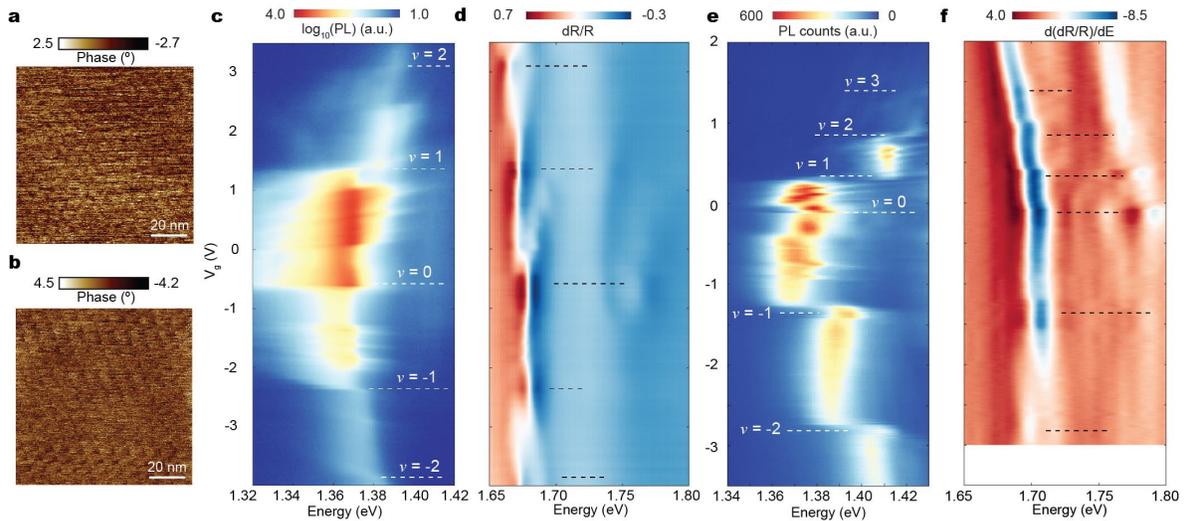

**Extended Data Figure 3. Moiré filling factor assignment. a-b,** PFM image of Device R1 (a) and Device H1 (b) described in the maintext. The moiré wavelengths are measured as 7.5 nm (R1) and 8 nm (H1). **c,** Gate-dependent interlayer exciton photoluminescence, taken at a different cool down compared to Fig. 2d in the maintext. The excitation energy is 1.96eV with the power 50 nW. Temperature is 4.7 K. **d,** corresponding differential optical reflectance spectra of Device R1. **c** and **d** share the same y axis. **e,** Gate-dependent interlayer exciton PL (same as Fig. 4c) of Device H1, and **f,** Corresponding differential optical reflectance spectra differentiated with respect to photon energy. **e** and **f** share the same y axis.

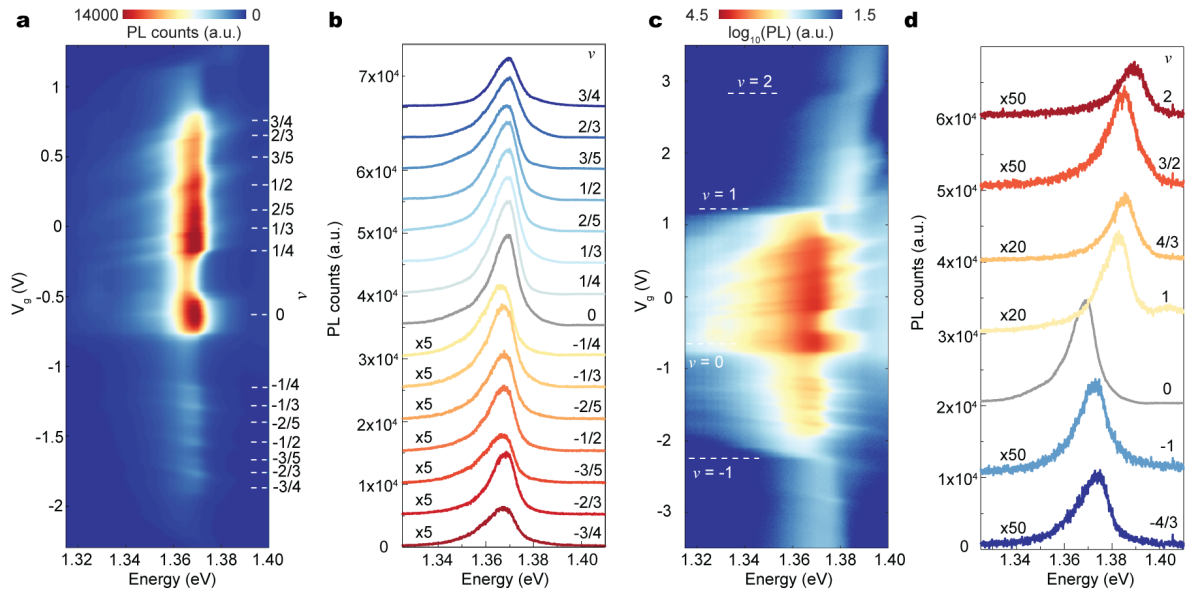

**Extended Data Figure 4. Interlayer exciton PL of Device R1 at selected filling factors.**
**a,** Zoom-in of gate-dependent interlayer exciton PL of Device R1 with |ν| < 1. **b,** Linecuts of PL spectra with fractional fillings. The PL spectra are evenly offset. At hole doping side, the PL counts are multiplied by 5 times. There is little variation of PL peak energy as ν varies. **c,** Gate-dependent interlayer exciton PL of Device R1, same as Fig. 2d in main text. **d,** Linecuts of PL spectra at integer filling conditions. The spectra are evenly offset. The PL counts are multiplied by either 20 or 50 times, as indicated in the plot, except at ν=0.

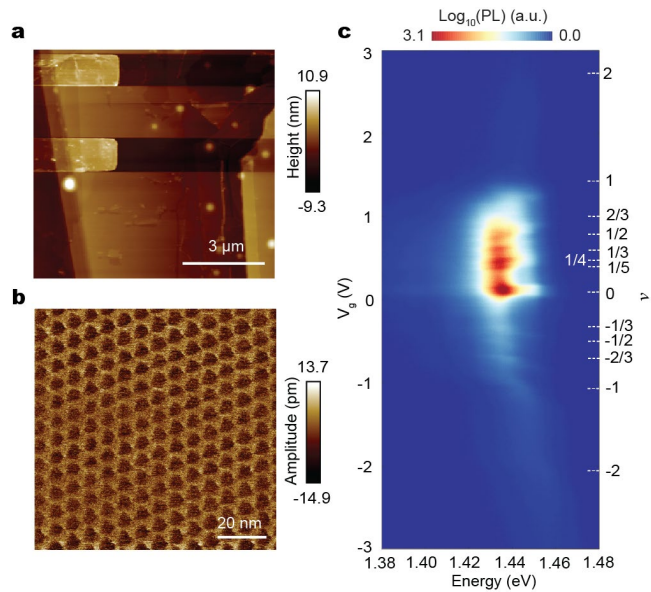

**Extended Data Figure 5. Additional R-stacked heterobilayer Device R2. a,** AFM morphology and **b,** PFM image of Device R2, showing the moiré wavelength is about 7.5 nm. **c.** Gate-dependent interlayer exciton photoluminescence of Device R2, with filling factors indicated.

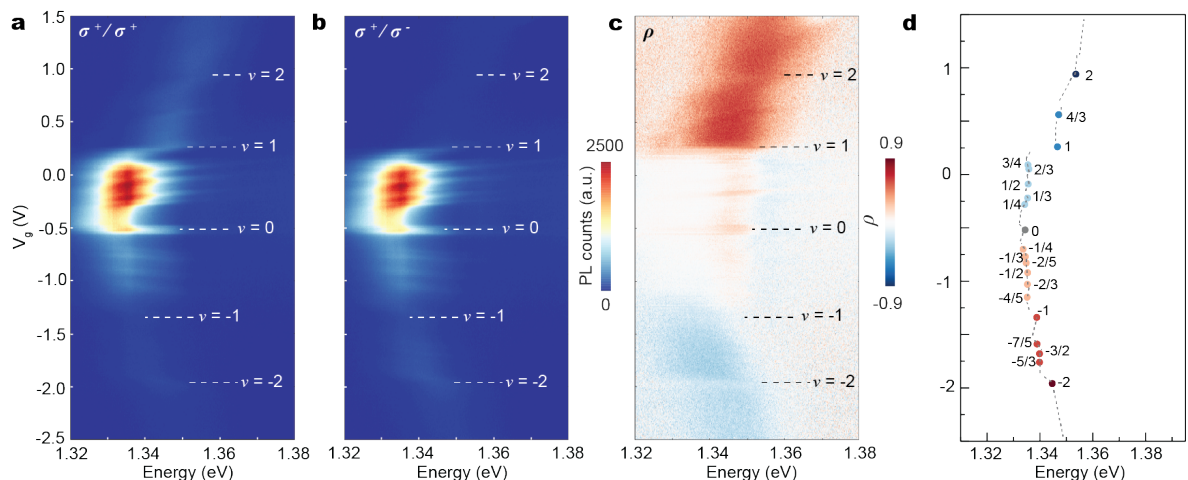

**Extended Data Figure 6. Polarization resolved PL of an additional R-stacked heterobilayer Device R3. a,** Co-circularly **b,** cross-circularly polarized interlayer exciton PL for Device-R3 under σ+ circularly polarized excitation. The excitation power is 60 nW with excitation energy 1.682 eV at 10 K. **c,** Corresponding degree of circularly polarization $\rho$. a-c share the same y axis. **d,** Extracted PL peak energies (dots) for each integer and fractionally filled correlated charge states. The dashed lines indicate PL peak energies at all measured gate voltages. Devices R3 and H2 (Extended Data Fig. 8) are different parts of the same sample.

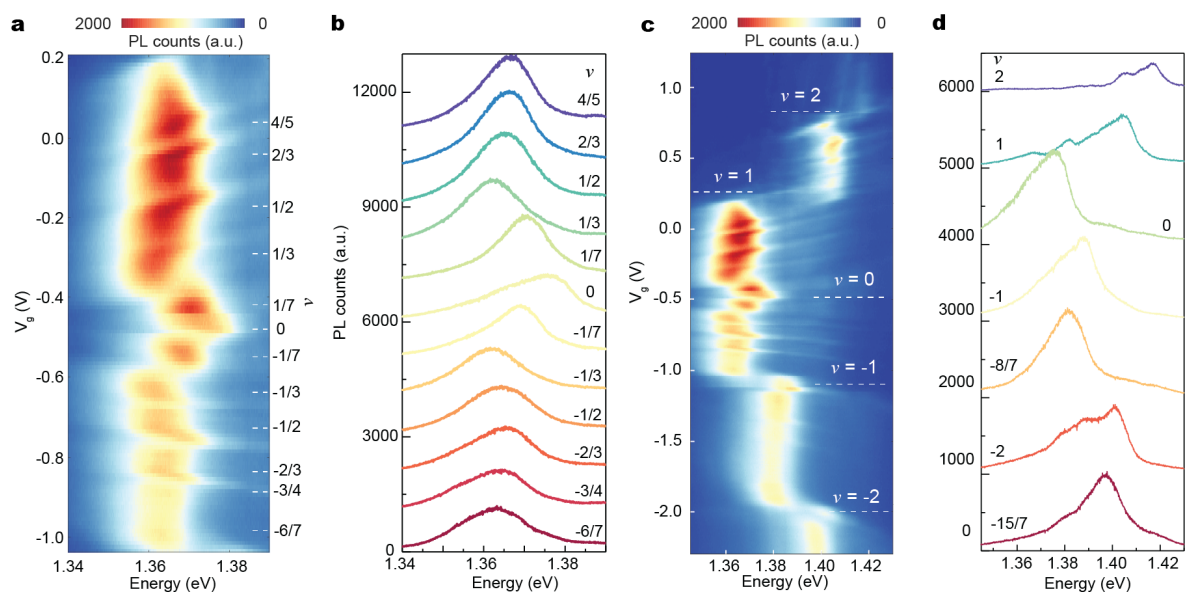

**Extended Data Figure 7. PL spectra of Device H1 at select filling factors. a,** Zoom-in of gate-dependent interlayer exciton PL with $|v| < 1$. **b,** Linecuts of PL spectra at fractional filling conditions as marked. **c,** Gate-dependent interlayer exciton PL, same as Fig. 2f in main text. **d,** Linecuts of PL spectra at integer and select fractional fillings. The spectra in **b** and **d** are evenly offset for clarity.

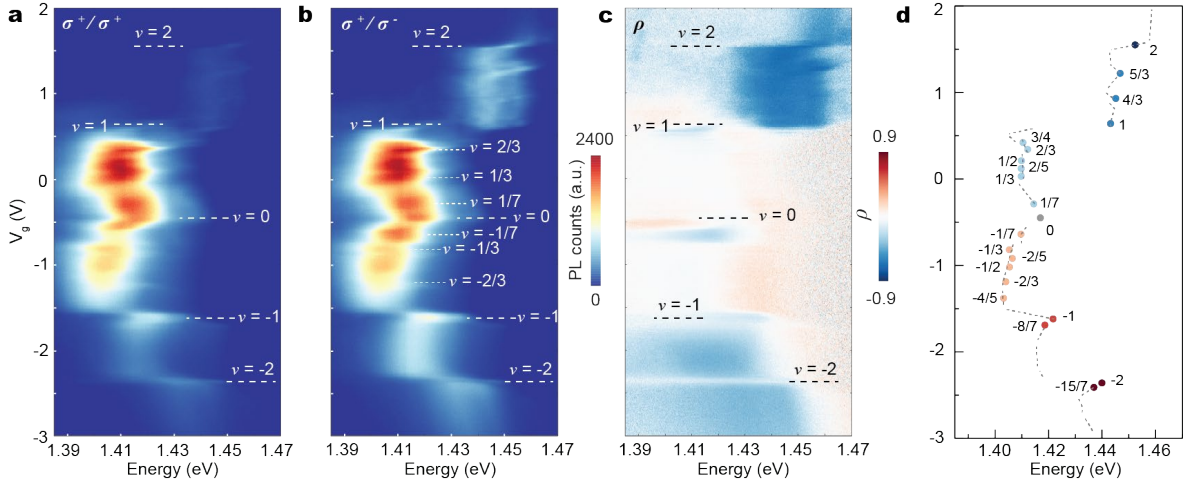

**Extended Data Figure 8. Polarization resolved PL of an additional H-stacked heterobilayer Device H2. a**, Co-circularly **b,** cross-circularly polarized interlayer exciton PL for Device-H2 under σ+ circularly polarized excitation. The excitation power is 60 nW with excitation energy 1.682 eV at 10 K. **c**, Corresponding degree of circularly polarization $\rho$. a-c share the same y axis. **d,** Extracted PL peak energies (dots) for each integer and fractionally filled correlated charge states. The dashed lines indicate PL peak energies at all measured gate voltages. Devices R3 (Extended Data Fig. 6) and H2 are different parts of the same sample.

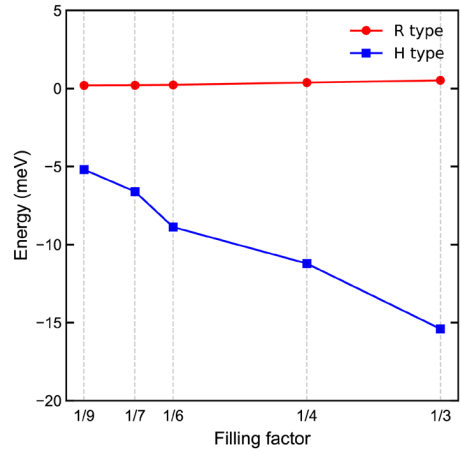

**Extended Data Figure 9. Calculated Coulomb interaction energy between exciton and electron-lattice for the filling factors ≤ 1/3.** Standard Ewald summation technique is used to calculate the Coulomb interaction energies for different filling factors. The convergence parameters have been carefully tested when doing the summation. The collective interaction is repulsive for R-stacking and attractive for H-stacking. Our model calculations match well for the two different interactions between exciton and charge lattices.

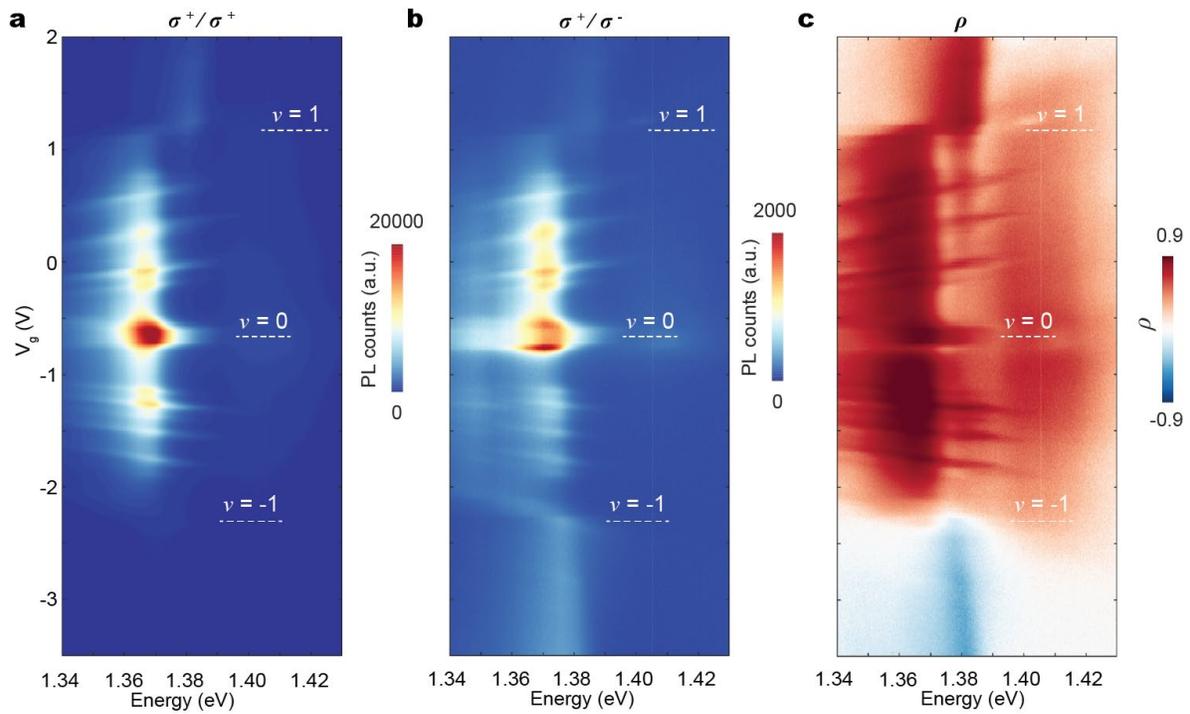

**Extended Data Figure 10. Polarization resolved PL of Device-R1 at a magnetic field of 8T. a,** Co-circularly and **b,** Cross-circularly polarized interlayer exciton PL under σ+ circularly polarized pump. **c,** The corresponding degree of circularly polarization ρ. Data is taken at 15 K. The optical excitation power is 300 nW with excitation energy at 1.678 eV.